\documentclass[%
 reprint,
superscriptaddress,
showpacs,
 amsmath,amssymb,
 aps,
floatfix,
]{revtex4-1}

\usepackage{graphicx}
\usepackage{dcolumn}
\usepackage{bm}

\usepackage{float}
\usepackage{amsfonts}
\usepackage{latexsym}
\usepackage{amssymb}
\usepackage{mathrsfs}
\usepackage{amsmath}
\usepackage{url} 
\usepackage{color}

\begin{document}

\preprint{APS/123-QED}
\title{An Information$-$Theoretic Framework to Measure the Dynamic Interaction between Neural Spike Trains}


\author{Gorana Mijatovic}
 \email{gorana86@uns.ac.rs}
\affiliation{%
 Faculty of Technical Sciences, University of Novi Sad, Serbia}%


\author{Yuri Antonacci}
\email{yuri.antonacci@unipa.it}
\affiliation{
 Department of Physics and Chemistry "Emilio Segrè", University of Palermo, Italy
}%

\author{Tatjana Loncar-Turukalo}
 \email{turukalo@uns.ac.rs}
\affiliation{%
 Faculty of Technical Sciences, University of Novi Sad, Serbia}%

\author{Ludovico Minati}
\email{lminati@ieee.org}
\affiliation{%
 Center for Mind/Brain Sciences (CIMeC), University of Trento, 38123 Trento, Italy, and the Institute of Innovative Research, Tokyo Institute of Technology, Yokohama 226-8503, Japan
}%

\author{Luca Faes}
\email{luca.faes@unipa.it}
\affiliation{%
 Department of Engineering, University of Palermo, Italy
}%


\date{\today}

\date{\today}

\begin{abstract}
Understanding the interaction patterns among simultaneous recordings of spike trains from multiple neuronal units is a key topic in neuroscience. However, an optimal approach of assessing these interactions has not been established, as existing methods either do not consider the inherent point process nature of spike trains or are based on parametric assumptions that may lead to wrong inferences if not met.
This work presents a framework, grounded in the field of information dynamics, for the model-free, continuous-time estimation of both undirected (symmetric) and directed (causal) interactions between pairs of spike trains.
The framework decomposes the overall information exchanged dynamically between two point processes $\textit{X}$ and $\textit{Y}$ as the sum of the dynamic mutual information (dMI) between the histories of $\textit{X}$ and $\textit{Y}$, plus the transfer entropy (TE) along the directions $\textit{X} \rightarrow \textit{Y}$ and $\textit{Y} \rightarrow \textit{X}$.
Building on recent work which derived theoretical expressions and consistent estimators for the TE in continuous time, we develop algorithms for estimating efficiently all measures in our framework through nearest neighbor statistics.
These algorithms are validated in simulations of independent and coupled spike train processes, showing the accuracy of dMI and TE in the assessment of undirected and directed interactions even for weakly coupled and short realizations, and proving the superiority of the continuous-time estimator over the discrete-time method.
Then, the usefulness of the framework is illustrated in a real data scenario of recordings from in-vitro preparations of spontaneously-growing cultures of cortical neurons, where we show the ability of dMI and TE to identify how the networks of undirected and directed spike train interactions change their topology through maturation of the neuronal cultures.
\end{abstract}

\maketitle


\section{\label{sec:level1}Introduction}
Multi-electrode recording techniques, which have become a standard in the neuroscience, provide large amounts of data about the neural activity measured at different temporal and spatial scales. In particular, simultaneous recordings of the firing activity of hundreds of neurons in various regions of the brain are nowadays widely available, and such availability is raising more and more the interest of neuroscientists about how groups of neurons influence reciprocally their firing  and act in concert to determine the function of a given brain region \cite{brown2004multiple}.

Thrust by this interest, the field of computational neuroscience has witnessed a continuous development of tools and algorithms to quantify the degree of interaction between two or more simultaneously recorded spike trains. The large body of work in this context has evolved along two main directions: the development of symmetric (undirected) measures of coupling between pairs of spike trains, commonly denoted as synchrony measures \cite{kreuz2007measuring}, and the design of causal (directed) interaction measures, typically based on the statistical concept of Granger causality (GC) \cite{granger1969investigating}. Synchrony measures are essentially aimed at detecting pairwise correlations in spike trains \cite{cutts2014detecting}, and are based on standard tools like cross-correlation \cite{eguiluz2005scale} and coherence \cite{salvador2005undirected} or on specific statistics applied to the inter-spike intervals (ISIs) extracted from the two analyzed trains \cite{kreuz2007measuring,kreuz2013monitoring}. On the other hand, directed measures attempt to infer the causal influence that one neural unit exerts over another by adopting different strategies for the implementation to spike train data of the concept of predictability improvement inherent in the definition of GC \cite{kim2011granger,nedungadi2009analyzing}. Up to now, these two approaches have been pursued as alternative to one another, and unifying frameworks embedding directed and undirected measures of spike train interaction as complementary aspects of the overall dynamic interaction between spike trains are lacking.

A major issue with the estimation of both undirected and directed interactions between neural spike trains is that the underlying tools usually do not consider the point process nature of neural spike train data.
In fact, interaction measures like correlation and GC are typically defined for continuously valued signals uniformly sampled in time (i.e., time series), and their computation for point processes poses both theoretical and practical challenges.
The application of these measures to neural spike trains has been performed transforming the spike train data either into discrete-time sequences through binning of the temporal axis (e.g., \cite{pasquale2008self}), or into continuously valued signals through convolution with smoothing kernels (e.g., \cite{zhu2003probing}). In either case, the alteration of the nature of the observed data causes a loss of information (in the case of time discretization) or the addition of distorting information (in the case of smoothing) which makes the estimation of the desired measure intrinsically sub-optimal. Moreover, the parameters associated with the transformation (e.g., width of the time bin, kernel type) are difficult to set and impact substantially on the estimated interaction measure. Alternatives which avoid data transformation have been proposed in the context of parametric modeling, for example in \cite{kim2011granger,valenza2017instantaneous} where linear parametric models were introduced to compute GC specificaly for point processes. However, parametric approaches are limited in the fact that they set \textit{a priori} the type of interaction to be studied, and may miss to reveal important aspects of such interaction if the data do not fit the assumed model. 

To overcome the above limitations, the present study introduces a framework for the computation of directed and undirected measures of interaction between neural spike trains, which is directly applicable to point processes evaluated in continuous time and does not require a model of the interactions. The framework is defined in the emerging field of information dynamics \cite{lizier2012local,faes2017information}, and exploits well-known concepts like mutual information (MI) \cite{cover1991elements} and transfer entropy (TE) \cite{schreiber2000measuring}. The MI is a model-free undirected measure of the 'static' interaction between two processes, while the TE is a dynamic and directed measure that implements the probabilistic notion of GC in a non-parametric fashion.
Here, we consider a measure of the overall information exchanged dynamically between two spike train processes $X$ and $Y$, and expand it as the sum of the dynamic MI (dMI) between the histories of $X$ and $Y$ and the TE along the two directions $X \rightarrow Y$ and $Y \rightarrow X$. Importantly, we build on the theoretical formulation of the TE for point processes \cite{spinney2017transfer} and on the recent introduction of a fully model-free estimator based on nearest neighbor statistics \cite{shorten2020estimating}, to compute all measures in our framework in continuous-time for spike train processes. The resulting dMI and TE measures are first validated in simulated scenarios of independent and coupled spike trains, and then used to identify the networks of undirected and directed interactions underlying the spiking activity of spontaneously-growing cultures of cortical neurons.

The Matlab Software relevant to this work is available for free download from the github repository https://github.com/mijatovicg/TEMI.

\section{Methods}
\subsection{Information-theoretic framework to assess the dynamic interaction between stochastic processes} \label{Sec_def}

\begin{figure*} [t]
    \centering
    \includegraphics[scale=0.75]{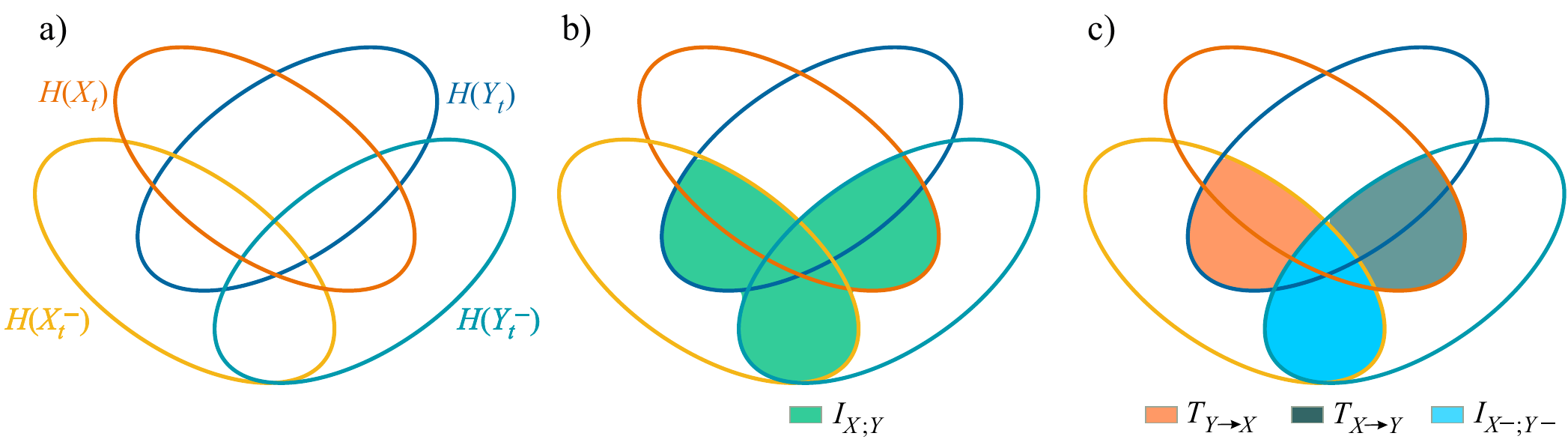}
    \caption{Venn diagram representation of the proposed measure of dynamic mutual information and of its decomposition. (a) Entropy of the present state ($H(X_t), H(Y_t)$) and of the past states ($H(X_t^-), H(Y_t^-)$) of two processes $X$ and $Y$ spike trains and the corresponding pasts  $X^-$ and $Y^-$. (b) Overall dynamic information shared between $X$ and $Y$, $I_{X;Y}$. Decomposition of $I_{X;Y}$ revealing the directional transfer entropies $T_{X\rightarrow Y}$ and $T_{Y\rightarrow X}$, and the symmetric dynamic mutual information $I_{X^-;Y^-}$.}
    \label{venn_diagram}
\end{figure*}

Given two possibly coupled dynamical systems $\mathcal{X}$ and $\mathcal{Y}$, we assume that their evolution over time is described by the stochastic processes $X=\{X_t\}$ and $Y=\{Y_t\}, t \in \mathbb{R}$. To evaluate the dynamic interaction between the two systems in the information-theoretic domain, we consider a measure quantifying the amount of information shared between the present and past states of the related processes. Formally, if $X_t, Y_t$ denote the present state of the two processes and $X_t^-,Y_t^-$ denote their past history, the overall dynamic information shared between $X$ and $Y$ is defined as:
\begin{equation}
\label{eq_dMI}    
I_{X;Y} = I(X_t^-;Y_t^-) + I(X_t;Y_t^-|X_t^-) + I(Y_t;X_t^-|Y_t^-),
\end{equation} 
where $I(\cdot;\cdot)$ and $I(\cdot;\cdot|\cdot)$ denote mutual information (MI) and conditional MI (CMI). In particular, the MI term in  (\ref{eq_dMI}) is a measure of the information shared by the past states of the two analyzed processes, which we denote as dynamic mutual information (dMI) between the histories of $X$ and $Y$,
$I_{X^-;Y^-} = I(X_t^-;Y_t^-)$; the two CMI terms correspond to the information transfer from $X$ to $Y$ and from $Y$ to $X$, $T_{X\rightarrow Y}=I(Y_t;X_t^-|Y_t^-)$ and $T_{Y\rightarrow X}=I(X_t;Y_t^-|X_t^-)$, quantified according to the well-known notion of transfer entropy (TE) \cite{schreiber2000measuring}. Note that in the definition of $I_{X^-;Y^-}$, $T_{X\rightarrow Y}$ and $T_{X\rightarrow Y}$ the time index is dropped under the hypothesis of stationarity of the processes.
Eq. (\ref{eq_dMI}) is explained graphically in the Venn diagram representation of Fig. \ref{venn_diagram}, and evidences how the overall dynamic interaction measure $I_{X;Y}$ is decomposed as the sum of the two directional TE measures, $T_{X\rightarrow Y}$ and $T_{Y\rightarrow X}$, plus the symmetric measure of dynamic MI, $I_{X^-;Y^-}$.

The information-theoretic measures defined above are typically computed for discrete-time stochastic processes, i.e. processes defined at discrete time instants $t_n=n\Delta t, n \in \mathbb{Z}$, where $\Delta t$ is the interval between time samples expressed in units of time.
In discrete time, information dynamic measures are well-established and a number of practical approaches exist to estimate dMI and TE measures starting from realizations of the processes $X$ and $Y$ provided in the form of synchronous time series of finite length \cite{wibral2014directed}.
The definition and subsequent computation of information dynamic measures in continuous time, i.e. in the limit $\Delta t \rightarrow 0$, is much more cumbersome and has been proposed only recently  \cite{spinney2017transfer,shorten2020estimating}.
In either case, to ensure convergence in the limit of small time bin size it is important to compute the information dynamic measures as \emph{rates}, normalizing them to the width of the time bins. 
Using information rates, the decomposition of the overall dynamic interaction between $X$ and $Y$ becomes: 
\begin{equation}
\label{eq_dMIrate}    
\dot{I}_{X;Y} = \dot{I}_{X^-;Y^-} + \dot{T}_{X\rightarrow Y} + \dot{T}_{Y\rightarrow X},
\end{equation} 
where the dMI and TE rates are obtained normalizing the corresponding terms in (\ref{eq_dMI}) by the time bin width (e.g., $\dot{I}_{X^-;Y^-}=(1/\Delta t) I_{X^-;Y^-}, \dot{T}_{Y\rightarrow X}=(1/\Delta t)T_{Y\rightarrow X}$).

In discrete-time, setting $t_n$ as the present time step, the random variables describing the present and past states of the processes $X$ and $Y$ are $X_{t_n}, Y_{t_n}$ and $X_{t_n}^-, Y_{t_n}^-$, respectively. With this notation, the dMI and TE rates appearing in (\ref{eq_dMIrate}) are formulated as follows:
\begin{equation}
\begin{split}
\label{eq_dMIdiscrete}
\dot{I}_{X^-;Y^-} & = \overline{\lambda} \cdot \mathbb{E}\left[ \ln \frac{p(x_n^-,y_n^-)}{p(x_n^-)p(y_n^-)} \right] \\
& = \lim_{T\to\infty} \frac{1}{T} \sum_{n=1}^{N} \ln \frac{p(x_n^-,y_n^-)}{p(x_n^-)p(y_n^-)},
\end{split}
\end{equation}

\begin{equation}
\begin{split}
\label{eq_TEdiscrete}
\dot{T}_{Y \rightarrow X} & = \overline{\lambda} \cdot \mathbb{E}\left[ \ln \frac{p(x_n|x_n^-,y_n^-)}{p(x_n|x_n^-)} \right] \\
& = \lim_{T\to\infty} \frac{1}{T} \sum_{n=1}^{N} \ln \frac{p(x_n|x_n^-,y_n^-)}{p(x_n|x_n^-)},
\end{split}
\end{equation}

\noindent{where $\overline{\lambda}=1/\Delta t = N/T$ is the sampling rate of the discrete-time processes, $T=N\Delta t$ is the duration of a time window containing $N$ samples of the processes, $x_n, y_n$ and $x_n^-=[x_{n-1},x_{n-2}\ldots], y_n^-=[y_{n-1},y_{n-2}\ldots]$, are realizations of $X_{t_n}, Y_{t_n}$ and of $X_{t_n}^-, Y_{t_n}^-$, and $p(\cdot)$, $p(\cdot,\cdot)$ and $p(\cdot|\cdot)$ denote marginal, joint and conditional probability density.} Note that the equivalences in (\ref{eq_dMIdiscrete}) and (\ref{eq_TEdiscrete}) hold for ergodic processes where time averages can replace ensemble averages. In the following section, we show how similar formulations can be obtained for a particular class of continuous-time processes, i.e., spike trains.

\subsection{Formulation of the framework for spike train processes} \label{Sec_framework}
In this section, the framework introduced above for the dynamic analysis of generic stochastic processes is formalized for spike train processes. Spike trains are continuous-time processes described by the occurrence at specific non-overlapping time instants of  indistinguishable events, or spikes. As such, spike trains are typically observed as series of time points corresponding to the event times; here, we consider the spike trains $X = \{x_{i}\}, i=1, 2, \dotsc, N_{X}$, and $Y = \{y_{j}\}, j=1, 2, \dotsc, N_{Y}$, where $x_{i}$ and $y_{j}$ are real numbers denoting the time points of the $i^{th}$ spike of the train $X$ and of the $j^{th}$ spike of the train $Y$. We stress that this is in contrast with the discrete-time formulation of stochastic processes, where realizations of the processes $X$ and $Y$ are time series of values measured at synchronously sampled time points.

While for discrete-time processes the system state is mapped by the time series values, in the case of point processes the state is defined at each time instant by the so-called counting process, i.e. by the continuous-time process that counts the number of spikes which have occurred up to the present time; for the spike train $X$, the counting process is $N_{X,t}=n$: $x_n \leq  t <x_{n+1}$ $(N_{X,t}=0$ $\forall t<x_1,  N_{X,t}=N_X$ $\forall t \geq X_{N_X})$. Then, the firing rate measures the probability for the train $X$ to fire in a time interval $[u, u+\Delta{u}]$ relative to the duration $\Delta{u}$ of the interval; given the counting process, the instantaneous firing rate computed at time $u$ for the process $X$ is $\lambda_{X,u} = \lim_{\Delta{u}\to{0}} p_u\big(N_{X,u+\Delta{u}} - N_{X,u} = 1 \big) / \Delta{u}$, where $p_u$ refers to a probability density evaluated in continuous time (i.e., at any time $u$).
With this formalism, the TE rate from the source spike train $Y$ to the target spike train $X$ is defined as \cite{spinney2017transfer}:

\begin{equation}
\begin{split}
\label{eq_TErate}
\dot{T}_{Y \rightarrow X}  & =  \overline{\lambda}_{X} \mathbb{E}_{p_{x}} \left[\ln{\dfrac{\lambda_{X,x_i|X_{x_i}^-, Y_{x_i}^-}}{\lambda_{X,x_i|X_{x_i}^-}}}\right] \\
 & = \lim_{T\to\infty} \dfrac{1}{T} \sum_{i=1}^{N_{X}} \ln{\dfrac{\lambda_{X,x_i|X_{x_i}^-,Y_{x_i}^-}}{\lambda_{X,x_i|X_{x_i}^-}}},
\end{split}
\end{equation}

\noindent{where $N_{X}$ is the number of spikes occurring in $X$ during the period $T$ and $\overline{\lambda}_{X} = N_{X}/T$ is the average firing rate of $X$.
In (\ref{eq_TErate}), $\lambda_{X,x_i|X_{x_i}^-}$ and $\lambda_{X,x_i|X_{x_i}^-,Y_{y_i}^-}$ are the instantaneous firing rates of the target process $X$ evaluated at the time of its $i^{th}$ spike $x_i$, respectively conditioned on the past history of the target process only and on the past histories of both the target and the source.}
Importantly, while the probability density $p_u$ defining the instantaneous firing rate is taken at any arbitrary time point (unconditionally to events in any process), the probabilities defining the conditional firing rates used to assess the TE rate are taken at the times of the spiking events in the target. This aspect, which is stressed making it explicit in (\ref{eq_TErate}) that the expectation is taken over the distribution $p_x$, is crucial for the further developments that lead to the estimation of the TE rate in continuous time. Indeed, expressing the conditional firing rates in terms of $p_u$, making a Bayes inversion and observing that $\lim_{\Delta{u}\to{0}} p_u(\cdot|N_{X,u+\Delta{u}} - N_{X,u} = 1 \big)=p_x(\cdot)$, (\ref{eq_TErate}) can be rewritten as \cite{shorten2020estimating}:   
\begin{equation}
\label{eq_TEdot}    
\dot{T}_{Y \rightarrow X} = \overline{\lambda}_{X}\mathbb{E}_{p_{x}} \left[\ln \left(\dfrac{p_{x}(X_{x_i}^-, Y_{x_i}^-)}{p_{u}(X_{x_i}^-, Y_{x_i}^-)} \cdot \dfrac{p_{u}(X_{x_i}^-)}{p_{x}(X_{x_i}^-)}\right)\right],
\end{equation}
which provides the basis for the estimation of the TE rate in continuous time reported in Sect. \ref{Sec_estimation}.

The TE rate from $X$ to $Y$ is defined in a straightforward way by inverting the role of source and target spike trains in the derivations above. As regards the dMI rate quantifying the symmetric interactions between the past histories of the two processes, we define it extending (\ref{eq_dMIdiscrete}) to spike train processes in analogy to how (\ref{eq_TEdiscrete}) was extended to yield (\ref{eq_TErate}). Specifically, given the joint and marginal probability densities of the past history of the two processes evaluated in continuous time, $p_{u}(X_{u}^-,Y_{u}^-)$, $p_{u}(X_{u}^-)$ and $p_{u}(Y_{u}^-)$, $u \in \mathbb{R}$, the dMI rate is defined as:
\begin{equation}
\label{eq_dMIdot}
\begin{split}
\dot{I}_{X^-;Y^-} & = \overline{\lambda}_{U}\mathbb{E}_{p_{u}} \left[\ln \dfrac{p_{u}(X_{u}^-, Y_{u}^-)}{p_{u}(X_{u}^-)p_{u}(Y_{u}^-)}\right] \\
 & = \lim_{T\to\infty} \dfrac{1}{T} \sum_{i=1}^{N_{U}} \ln \dfrac{p_{u}(X_{u_i}^-, Y_{u_i}^-)}{p_{u}(X_{u_i}^-)p_{u}(Y_{u_i}^-)},
 \end{split}
\end{equation}
where $N_{U}$ time points, $u_1,\ldots,u_{N_U}$, are assumed to occur with intensity $\overline{\lambda}_{U} = N_{U}/T$ during a period of duration $T$. Similarly to (\ref{eq_TEdot}), the derivation in (\ref{eq_dMIdot}) provides a means to estimate the dMI rate for spike trains.

\subsection{Continuous time estimation}
\label{Sec_estimation}
This section presents the estimation of the TE and dMI measures composing the overall dynamic information shared by two spike trains $X$ and $Y$. The measures are introduced in (\ref{eq_dMI}) for generic stochastic processes and are made explicit as rates in (\ref{eq_TEdot}) and (\ref{eq_dMIdot}) for the case of spike train processes. The estimation approach is based first on building realizations of the past of the two processes from the available data by means of a history embedding strategy \cite{shorten2020estimating}, and then on applying on such realizations the nearest neighbor entropy estimator \cite{kozachenko1987sample} to compute the different entropy terms that compose the TE and dMI rates to be estimated; entropy estimation is performed following a strategy that favors compensation of the bias of the individual entropy terms when they are summed to get the desired measure \cite{kraskov2004estimating,shorten2020estimating}.

History embedding is performed in order to approximate the past of the two spike trains referred to specific time points such as the spike times $x_i$ or $y_j$, or to arbitrary time points $u_i$ sampled in continuous time. In the first case, the histories needed to compute the TE rate (\ref{eq_TEdot}) are constructed, from the set of target spike times $x_i, i=1,\ldots,N_X$, as illustrated in Fig. 2a: the history of the target train $X$ referred to the $i^{th}$ spike time $x_i$ is approximated taking $l$ inter-spike intervals as $X_{x_i}^- \approx X_{x_i}^l=[X_{x_i,1}, \cdots, X_{x_i,l}]$, where $X_{x_i,k}=x_{i-k+1}-x_{i-k}, k=1,\ldots l$; the history of the driver train $Y$ referred to $x_i$ is approximated taking first the interval from the most recent driver spike (i.e., $y_p: y_p<x_i, y_{p+1}\geq x_i$) to $x_i$ and then the preceding $l-1$ ISIs, i.e.  $Y_{x_i}^- \approx Y_{x_i}^l=[x_i-y_p,Y_{y_p}^{l-1}]$; the joint history at $x_i$, $J_{x_i}^-=[X_{x_i}^-,Y_{x_i}^-]$ is approximated by the $2l$ vector $J_{x_i}^l=[X_{x_i}^l,Y_{x_i}^l]$. In the second case, the histories needed to compute the MI rate (\ref{eq_dMIdot}) are constructed, from a set of arbitrary times $u_i, i=1,\ldots,N_U$, as illustrated in Fig. 2b: in this case the histories of both spike trains referred to $u_i$ are approximated taking the interval from the most recent spike to $u_i$ followed by $l-1$ ISIs, i.e. $X_{u_i}^- \approx X_{u_i}^l=[u_i-x_p,X_{x_p}^{l-1}]$, $Y_{u_i}^- \approx Y_{u_i}^l=[u_i-y_p,Y_{y_p}^{l-1}]$, $J_{u_i}^- \approx J_{u_i}^l=[X_{u_i}^l,Y_{u_i}^l]$; the times $u_i$ are placed randomly over the time axis, according to criteria determined depending on the application. Note that here we assume the same length $l$ for all history embeddings, but this can be optimized for each embedding separately \cite{shorten2020estimating}.

\begin{figure} [t!]
    \centering
    \includegraphics[scale=0.85]{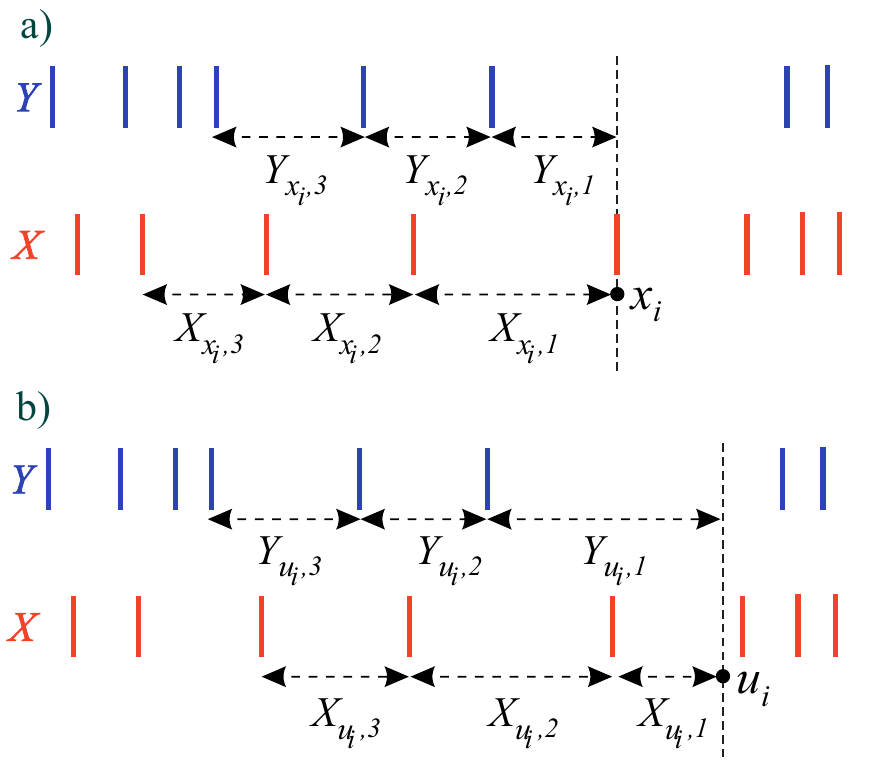}
    \caption{Example of history embeddings reconstructed with embedding length $l=3$. (a) Joint history embeddings $J_{x_i}^l=[X_{x_i},Y_{x_i}]$ at a target event $x_{i}$. (b) Joint history embeddings $J_{u_i}^l=[X_{u_i},Y_{u_i}]$ at a random event $u_{i}$.}
    \label{embeddings}
\end{figure}

The history embeddings built as described above at target spike times form the data matrices $\mathbf{X}^l_x \in \mathbb{R}^{(N'_X \times l)}$ and $\mathbf{J}^l_x \in \mathbb{R}^{(N'_X \times 2l)}$ which contain in the $n^{th}$ row respectively the vectors $X_{x_{n+l}}^l$ and $J_{x_{n+l}}^l$, $n=1,\ldots,N'_X$, while the history embeddings at arbitrary times form the data matrices $\mathbf{X}^l_u \in \mathbb{R}^{(N_U' \times l)}$ and $\mathbf{J}^l_u \in \mathbb{R}^{(N_U' \times 2l)}$ which contain in the $n^{th}$ row respectively the vectors $X_{u_{n}}^l$ and $J_{u_{n}}^l$, $n=1,\ldots,N_U'$ ($N'_X$ and $N_U'$ are in general lower than $N_X$ and $N_U$, see supplementary material).

These data matrices are passed as input to the algorithms for the estimation of TE and dMI rates, whose pseudo-codes and codes are provided in the supplementary material. 
Estimation of the TE rate and of the dMI rate is performed by first expanding (\ref{eq_TEdot}) and (\ref{eq_dMIdot}) in differential entropy terms, to obtain the equations $\dot{T}_{Y \rightarrow X}=\overline{\lambda}_{X}T_{Y\rightarrow X}$ and $\dot{I}_{X^-; Y^-}= \overline{\lambda}_{U}I_{X^-; Y^-}$, where
\begin{equation}
\label{est_TEdot}    
T_{Y\rightarrow X} = H_{p_u}(X_{x_i}^-, Y_{x_i}^-)- H(X_{x_i}^-, Y_{x_i}^-)  + H(X_{x_i}^-) - H_{p_u}(X_{x_i}^-)
\end{equation}
\begin{equation}
\label{est_dMIdot}    
I_{X^-; Y^-} = H(X_{u_i}^-) + H(Y_{u_i}^-)-H(X_{u_i}^-, Y_{u_i}^-);
\end{equation} 
the notations $H(\cdot)$ and $H_{p_u}(\cdot)$ refer to 'standard' differential entropy where expectation is taken over the same probability distribution for which the log-likelihood is estimated (e.g., $H(X_{x_i}^-)=-\mathbb{E}_{p_x}[\ln p_x(X_{x_i}^-)]$), and to 'cross-entropy' where the two distributions differ
(e.g., $H_{p_u}(X_{x_i}^-)=-\mathbb{E}_{p_x}[\ln p_u(X_{x_i}^-)]$) \cite{shorten2020estimating}.
Then, each entropy term is estimated using the well-known Kozachenko-Leonenko (KL) method \cite{kozachenko1987sample}. Starting from $N$ realizations of a generic $d$-dimensional variable $W$ forming the data matrix $\mathbf{W}\in \mathbb{R}^{(N \times d)}$, this method estimates the differential entropy $H(W)=-\mathbb{E}_{p_w}[\ln p_w(w)]$ as:
\begin{equation}
\label{HKL}    
\hat{H}(W) = \ln(N-1) -\psi(k) + \ln c_{d} + \dfrac{d}{N} \sum_{i=1}^{N} \ln \epsilon_{w_i,k,\mathbf{W}}
\end{equation}
where $\psi(\cdot)$ is the digamma function, $c_d$ is the volume of the $d$-dimensional unit ball under a given norm ($c_d=1$ for the maximum norm used in this work), and $\epsilon_{w_i,k,\mathbf{W}}$ is twice the distance between the $i^{th}$ realization of $W$ and its $k^{th}$ nearest neighbor taken from $\mathbf{W}$. To estimate a cross-entropy term $H_{p_v}(W)=-\mathbb{E}_{p_w}[\ln p_v(w)]$ from two data matrices $\mathbf{W}\in \mathbb{R}^{(N \times d)}$ and $\mathbf{V}\in \mathbb{R}^{(M \times d)}$, eq. (\ref{HKL}) is modified as
\begin{equation}
\label{HKLcross}    
\hat{H}_{p_v}(W) = \ln(M) -\psi(k) + \ln c_{d} + \dfrac{d}{N} \sum_{i=1}^{N} \ln \epsilon_{w_i,k,\mathbf{V}}
\end{equation}
where $\epsilon_{w_i,k,\mathbf{V}}$ is twice the distance between the vector $w_i \in \mathbf{W}$ and its $k^{th}$ nearest neighbor taken from $\mathbf{V}$. 

The formulations of the differential entropy and cross-entropy are exploited to compute the four entropy terms composing the TE in (\ref{est_TEdot}) and the three entropy terms composing the dMI in (\ref{est_dMIdot}), using from time to time the data matrices resulting from history embedding in place of the generic matrices $\mathbf{W}$ and $\mathbf{V}$.
While a naive estimator would fix the parameter $k$ and use (\ref{HKL}) and (\ref{HKLcross}) for each entropy term composing $\dot{T}_{Y \rightarrow X}$ and $\dot{I}_{X^-, Y^-}$, here we adopt the bias compensation strategies proposed in \cite{shorten2020estimating,faes2015estimating}, whereby the number of neighbors $k$ is changed at each data sample in order to use the same range of distances in spaces of different dimensions and ultimately reduce the bias in the estimation of sums of entropies.
The two strategies are implemented in Algorithm 1 and Algorithm 2 described in the supplementary material. Both algorithms start with a fixed parameter $k_{global}$ that will be the minimum number of nearest neighbors in any search space. At the $i^{th}$ iteration, corresponding to a realization $w_i$ of the data matrix $\textbf{W}$, the algorithms perform 
both neighbor searches whereby $k$ is fixed and the distance $\epsilon_{w_i,k,\mathbf{W}}$ between $w_i$ and its $k^{th}$ nearest neighbor within $\textbf{W}$ (or within a different data matrix $\textbf{V}$ in case of cross-entropy estimation) is computed, and range searches whereby the number of neighbors $k_{w_i,\mathbf{W}}$ of $w_i$ found inside $\textbf{W}$ (or inside $\textbf{V}$ in case of cross-entropy) is counted. Applying (\ref{est_TEdot}) and (\ref{est_dMIdot}) with entropy and cross-entropy as in (\ref{HKL}) and (\ref{HKLcross}) where the role of $w_i$ is taken by $X^l_{x_i}$, $J^l_{x_i}$, $X^l_{u_i}$, $Y^l_{u_i}$ or $J^l_{u_i}$, leads finally to estimate the TE rate as the output of Algorithm 1 \cite{shorten2020estimating}, and the dMI rate as the output of Algorithm 2 \cite{faes2015estimating}, as follows:

\begin{widetext}
\begin{equation}
\label{eq18}   
\dot{T}_{Y \rightarrow X} = \dfrac{\overline{\lambda}_X}{N'_X} \sum_{i=1}^{N'_{X}}   \psi(k_{X^l_{x_i},\mathbf{X}^l_u}) -\psi(k_{X^l_{x_i},\mathbf{X}^l_x}) + \psi( k_{J^l_{x_i}, \mathbf{J}^l_x})-\psi(k_{J^l_{x_i},\mathbf{J}^l_u})
 +  l \ln \dfrac{\epsilon_{X^l_{x_i},k_{X^l_{x_i},\mathbf{X}^l_x},\mathbf{X}^l_x} \cdot \epsilon^2_{J^l_{x_i},k_{J^l_{x_i},\mathbf{J}^l_x},\mathbf{J}^l_x}}{\epsilon_{X^l_{x_i},k_{X^l_{x_i},\mathbf{X}^l_u},\mathbf{X}^l_u} \cdot \epsilon^2_{J^l_{x_i},k_{J^l_{x_i},\mathbf{J}^l_u},\mathbf{J}^l_u}}, 
\end{equation}
 
\begin{equation}
\label{eq19} 
\dot{I}_{X^-;Y^-} = \overline{\lambda}_U \big[ \psi(k_{global}) + \ln (N'_U-1) - \dfrac{1}{N'_U} \sum_{i=1}^{N'_{U}} \big(
 \psi(k_{X^l_{u_i},\mathbf{X}^l_u}) + \psi(k_{Y^l_{u_i},\mathbf{Y}^l_u}) \big) \big] .
\end{equation}
\end{widetext}

\section{Validation on simulated spike trains}

This Section reports the application of the proposed analysis framework
on synthetic spike trains simulated under controlled conditions of firing and synchrony. Two different simulation scenarios are designed to reproduce the spiking dynamics of uncoupled processes and of coupled processes with different direction and intensity of interaction. In both simulations, the continuous-time measures of dMI rate and TE rate defined in Sect. \ref{Sec_framework} and estimated as reported in Sect. \ref{Sec_estimation} are compared with the discrete-time estimates of the same measures; the latter are obtained dividing the temporal axis into time bins, building the discrete-time process that 
counts the number of spikes falling into each time bin,
and applying Eqs. (\ref{eq_dMIdiscrete}) and (\ref{eq_TEdiscrete}) to the resulting 
sequences of natural numbers.

Both simulations are relevant to Poisson spike trains with mean firing rate $\overline{\lambda}=1$ spike/s. The first scenario considers pairs of independent Poisson trains $X$ and $Y$, for which the ground truth value of the overall dynamic information exchanged between the processes is zero ($\dot{I}_{X^-;Y^-}=\dot{T}_{X \rightarrow Y}=\dot{T}_{Y \rightarrow X}=0$). Process realizations are generated with variable duration, to simulate a number of target events $N_X=N_Y \in \{ 100,300, 500, 1000\}$ spikes.
The second scenario examines coupled Poisson spike trains which can be coupled unidirectionally or without a preferential direction of interaction. This is achieved generating a master process $X$ as a Poisson spike train, and a driven process $Y$ such that, for each spike occurring at time $x_i$ in $X$, a spike in $Y$ is placed at the time point $y_i = x_i + \tau + u_i$, where $\tau$ is a constant time delay and $u_i$ is a random time jitter sampled from the uniform distribution $\mathcal{U}(-\delta, \delta)$. With this setting, the parameter $\delta$ is inversely related to the coupling between the two trains, and changing the delay $\tau$ it is possible to achieve different coupling 
configurations:
$\tau=0$ corresponds to absence of a preferential coupling direction, while $\tau<0$ and $\tau>0$ determine unidirectional coupling, respectively from $Y$ to $X$ and from $X$ to $Y$.
Here we investigate the behavior of the dMI and TE rates when $\tau = a\delta$, with $a \in\{ -1, -0.5, 0, 0.5, 1\}$, and 
reproducing conditions from very strong coupling ($\delta \approx$ msec) to very weak coupling ($\delta \approx$ s).

In all simulations, the continuous time estimator is implemented choosing a number of arbitrary time points for entropy estimation equal to the number of simulated spikes $N_U=N_X=N_Y$, and drawing such points from the uniform distribution $\mathcal{U}(0,T)$; the length of the history embedding and the initial parameter for nearest neighbor analysis were set as $l=1$ and $k_{global}=5$.
The discrete-time estimator is implemented using combinations of the time bin width and embedding length set to cover the average duration of the inter-spike intervals ($\Delta t \cdot l=1$ s, $\Delta t=0.1$ s). Further evaluation of the dependence of the dMI and TE rates on the analysis parameters are reported in the supplementary material (Figs. S1, S2).

In each simulation, the ability of the continuous and discrete-time estimators to reveal the absence or presence of a connection is tested assessing the statistical significance of the estimated dMI and TE values. To this end, for any given measure, the value estimated for the pair of simulated spike trains is compared with the distribution of values of the same measure obtained from 100 pairs of surrogate spike trains generated under the null hypothesis of uncoupling. Surrogate trains were produced using the JOint DIstribution of successive inter-event intervals (JODI) algorithm \cite{ricci2019generation}, which retains the amplitude distribution and approximates the auto-correlation of the inter-spike intervals of the two original trains while destroying any coupling between them. According to a one-tailed hypothesis test with $5\%$ significance, the investigated measure ($\dot{T}_{Y \rightarrow X}$, $\dot{T}_{X \rightarrow Y}$, or $\dot{I}_{X^-, Y^-}$) is deemed as statistically significant if its value on the original trains exceeds the $95^{th}$ percentile of its distribution on surrogates.

\begin{figure} [t!]
    \centering
    \includegraphics[scale=0.8]{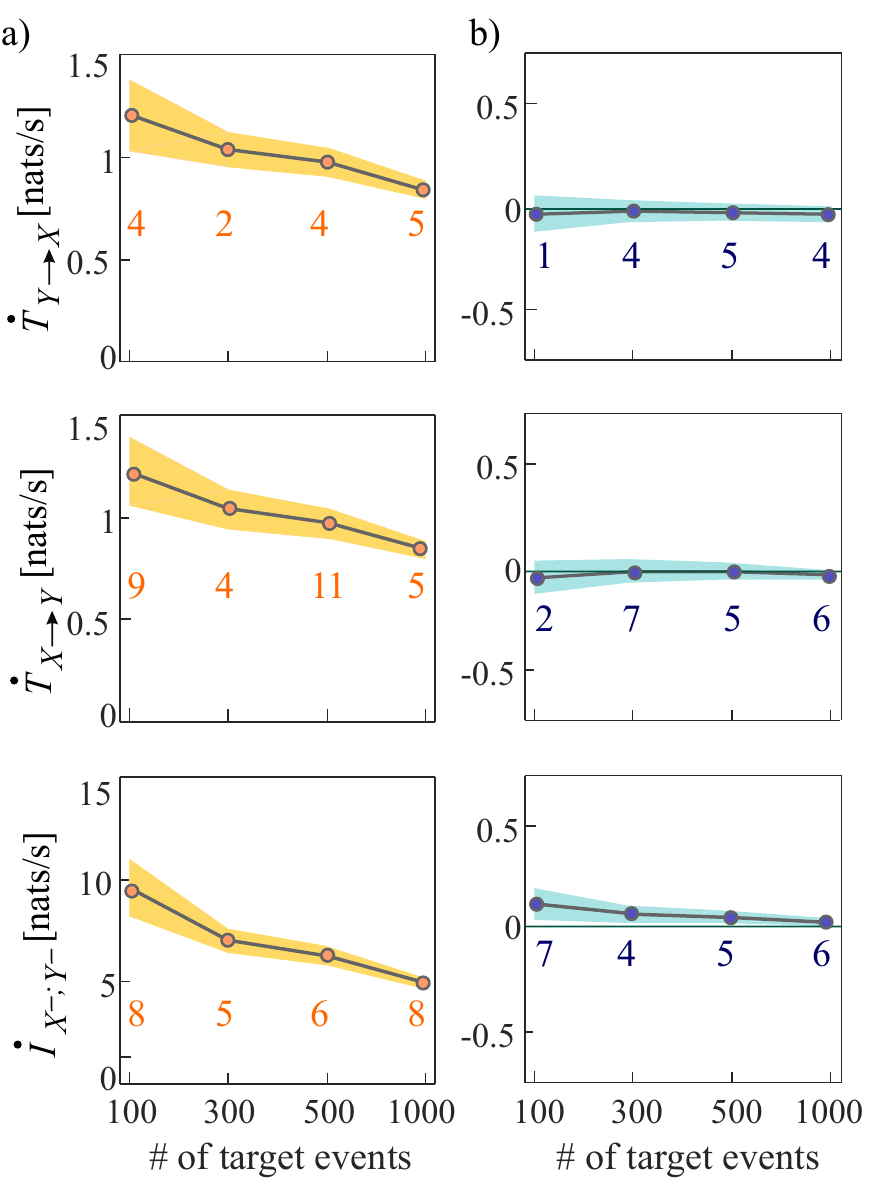}
    \caption{Comparison between discrete-time (a, orange) and continuous-time (b, blue) estimators of the TE and dMI rates applied to independent Poisson spike trains. Plots depict the mean (circles) and standard deviation (shades) of each measure, computed over 100 simulation runs as a function of the spike train duration. The number of realizations detected as statistically significant using JODI surrogates is reported for each measure and duration.}
    \label{FigSimuIndCmp}
\end{figure}

The comparison between the performances of continuous- and discrete-time estimators on independent processes is illustrated in Fig. \ref{FigSimuIndCmp}. For each measure, the performance can be inferred in terms of bias (i.e, the deviation of the average value across realizations from the expected zero level) and variance across realizations.
The discrete-time estimates of both TE and dMI exhibit a substantial bias for all the reported data lengths, while the continuous-time estimates are always very close to the true value. The variance is also lower for the continuous-time estimates, especially as regards the dMI measure. As expected, both estimators improve their performance with increasing the data length. As regards the statistical significance of the detected interactions, both estimators rejected the null hypothesis of uncoupling in a limited number of realizations, compatible with the nominal rate of false positives.

\begin{figure} [t!]
    \centering
    \includegraphics[scale=0.8]{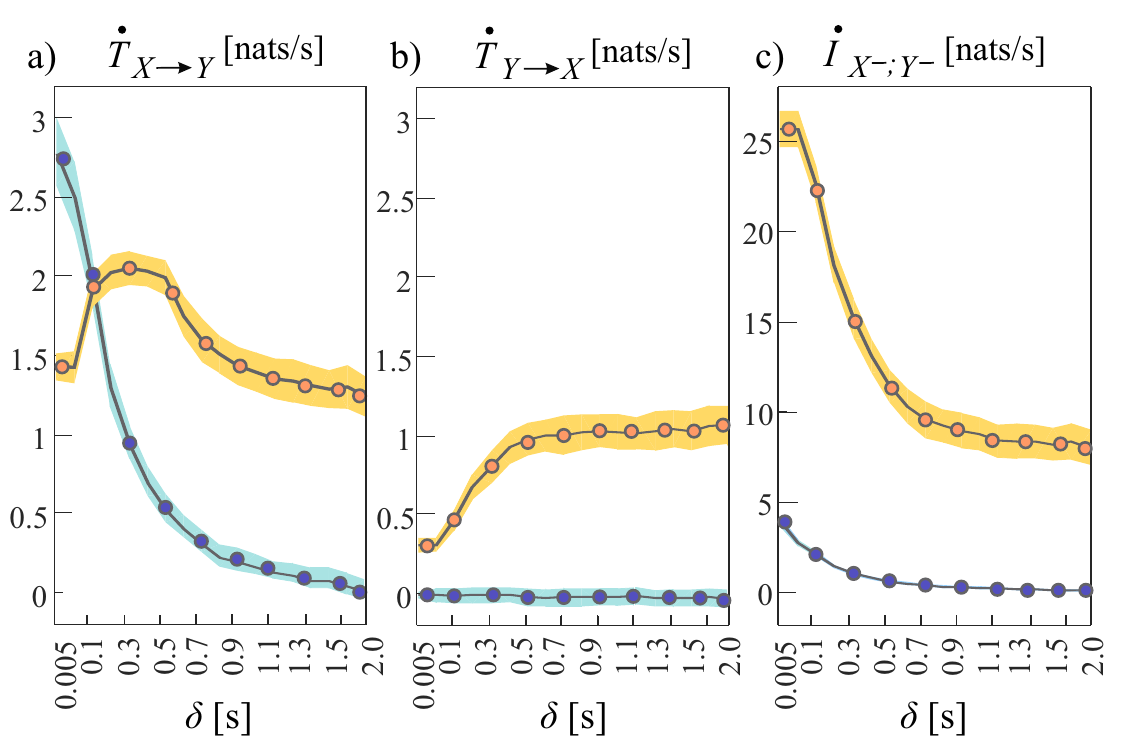}
    \caption{Comparison between discrete-time (orange) and continuous-time (blue) estimators of the TE and dMI rates applied to interacting Poisson spike trains coupled from $X$ to $Y$. Plots depict the mean (circles) and standard deviation (shades) of each measure, computed over 100 simulation runs as a function of the de-coupling parameter $\delta$.}
    \label{FigSimuDipCmp}
\end{figure}

In Fig. \ref{FigSimuDipCmp} the two estimators are compared for
pairs of spike trains of length $T=300$ s, interacting along the direction $X \rightarrow Y$ ($\tau=\delta$) with varying coupling strength modulated inversely by the parameter $\delta$.
The progressive de-coupling of the two trains obtained increasing $\delta$ is reflected by the TE rate from $X$ to $Y$ estimated in continuous time, while the discrete-time estimates exhibit a less interpretable non-monotonic behavior (Fig. \ref{FigSimuDipCmp}a). Along the uncoupled direction, the discrete-time TE rates deviate substantially from the expected zero level and display a bias dependent on $\delta$, while the continuous-time estimates are consistently null (Fig. \ref{FigSimuDipCmp}b). The MI rate decreases with $\delta$ for both estimators, but the discrete-time estimates seem to be again biased as they stabilize at $\sim 8$ nats/s when $\delta$ reaches the highest values corresponding to maximum de-coupling  (Fig. \ref{FigSimuDipCmp}c).

\begin{figure} [t!]
    \centering
    \includegraphics[scale=0.87]{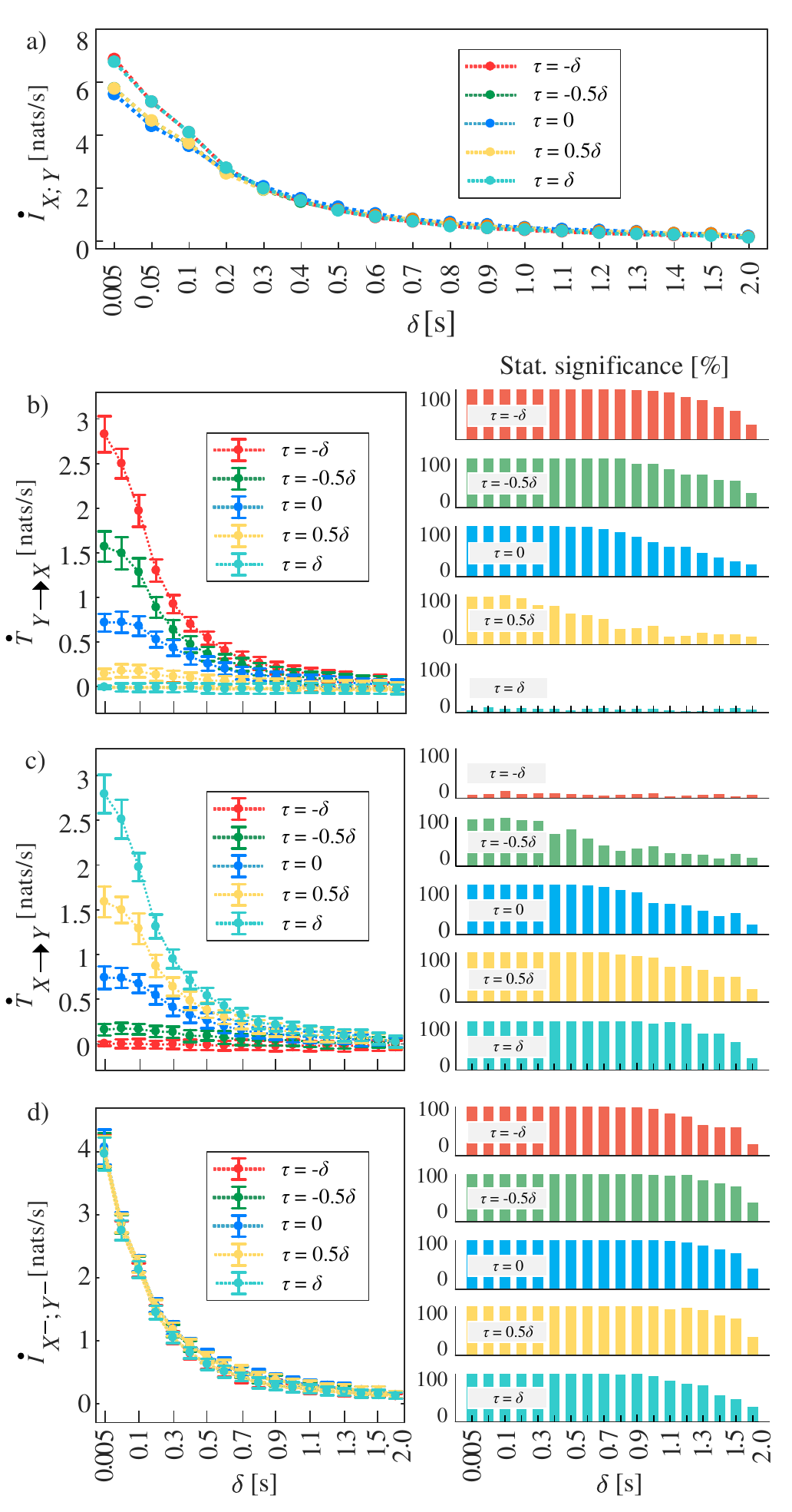}
    \caption{Continuous-time estimation the overall dynamic information shared between coupled Poisson spike trains (a) and of its TE rate (b,c) and dMI rate (d) components. Plots depict the distribution (mean $\pm$ SD) of each measure, computed over 100 simulation runs as a function of the de-coupling parameter $\delta$, for different values of the parameter $\tau$ determining the coupling direction. Bar plots report the number of realizations for which the considered measure is detected as statistically significant using JODI surrogates.}
    \label{coupled_Poisson}
\end{figure}

Fig. \ref{coupled_Poisson} provides an exhaustive description of the scenario with coupled Poisson spike trains investigated by the proposed continuous-time estimator. Simulations of length $T=300$ s are iterated to reproduce conditions of strong ($\tau=\pm \delta$) and intermediate ($\tau=\pm 0.5\delta$) coupling from $X$ to $Y$ ($\tau>0$) and from $Y$ to $X$ ($\tau<0$), as well as bidirectional coupling ($\tau=0$), each with coupling strength decreasing progressively as the de-coupling parameter increases from $\delta=0.005$ to $\delta=2$.
The overall dynamic information exchanged between the two trains ($\dot{I}_{X;Y}$ , Fig. \ref{coupled_Poisson}a) is insensitive to the coupling direction, as seen by the overlap of its average values for $\tau=\pm \delta$ and for $\tau=\pm 0.5\delta$, and reflects the coupling strength, as seen by its monotonic decrease observed at increasing $\delta$.
The TE rates along the two directions of interaction between $X$ and $Y$ (Fig. \ref{coupled_Poisson}b,c) exhibit symmetric trends, with $\dot{T}_{Y \rightarrow X}$ decreasing progressively and $\dot{T}_{X \rightarrow Y}$ increasing progressively as the preferential coupling direction shifts from $Y \rightarrow X$ ($\tau=-\delta$) to from $X \rightarrow Y$ ($\tau=\delta$). Both TE rates decrease with increasing the de-coupling parameter and start losing statistical significance for $\delta$ higher than $\simeq 0.9$, becoming largely non-significant for $\delta=2$; the analysis of surrogate data indicates also the absence of directed interactions along the uncoupled direction, as documented by the small rate of detection of significant $\dot{T}_{Y \rightarrow X}$ values when $\tau=\delta$ and of significant $\dot{T}_{Y \rightarrow X}$ values when $\tau=-\delta$.
Finally, the rate of dynamic information shared by the process histories ($\dot{I}_{X^-;Y^-}$ , Fig. \ref{coupled_Poisson}d) decreases monotonically with $\delta$ and is almost insensitive to $\tau$; the dMI rate remains high and statistically significant also when the coupling direction is not unequivocal ($\tau=\pm0.5\delta$) or not established ($\tau=0$), providing in these conditions higher detection rates than the TE even when the coupling is weak ($\delta \geq 1$).

In sum, the reported simulations demonstrate the consistency and accuracy of the continuous-time estimator of the rate of dynamic information shared between two spike train processes and of its dMI and TE components, as well as its superiority over the standard discrete-time estimation. Importantly, the TE rate estimator is able to detect directed interactions even for weakly coupled and short spike trains, and the dMI rate estimator is able to detect undirected interactions in challenging conditions of weak bidirectional coupling.

\section{Application}
To test the usefulness of the presented framework in a real-world scenario of interacting neural spike trains, we considered the data from a public repository of in-vitro cultures of dissociated cells grown on multi-electrode arrays (MEAs) \cite{morin2005investigating,wagenaar2006extremely}. In this experimental setting, neocortical cells were harvested from the brains of rat embryos and plated on glass culture wells; each culture was obtained plating $\sim$ 50000 cells in a droplet instrumented by a MEA with a grid of $8 \times 8$ electrodes, in which each electrode captures the spontaneous firing activity of $\sim$ 100-1000 neurons \cite{morin2005investigating}.
This recording modality achieves a mesoscale spatial description of the neural activity. A similar mesoscale representation is obtained also in time, due to the temporal binning of the spike times by the sampling rate of the analog-to-digital (ADC) converter; this temporal coarse graining does not detract in any way the opportunity to use our continuous time measures, as the spiking nature of the data remains intact.

\begin{figure*} [t]
    \centering
    \includegraphics[scale=0.9]{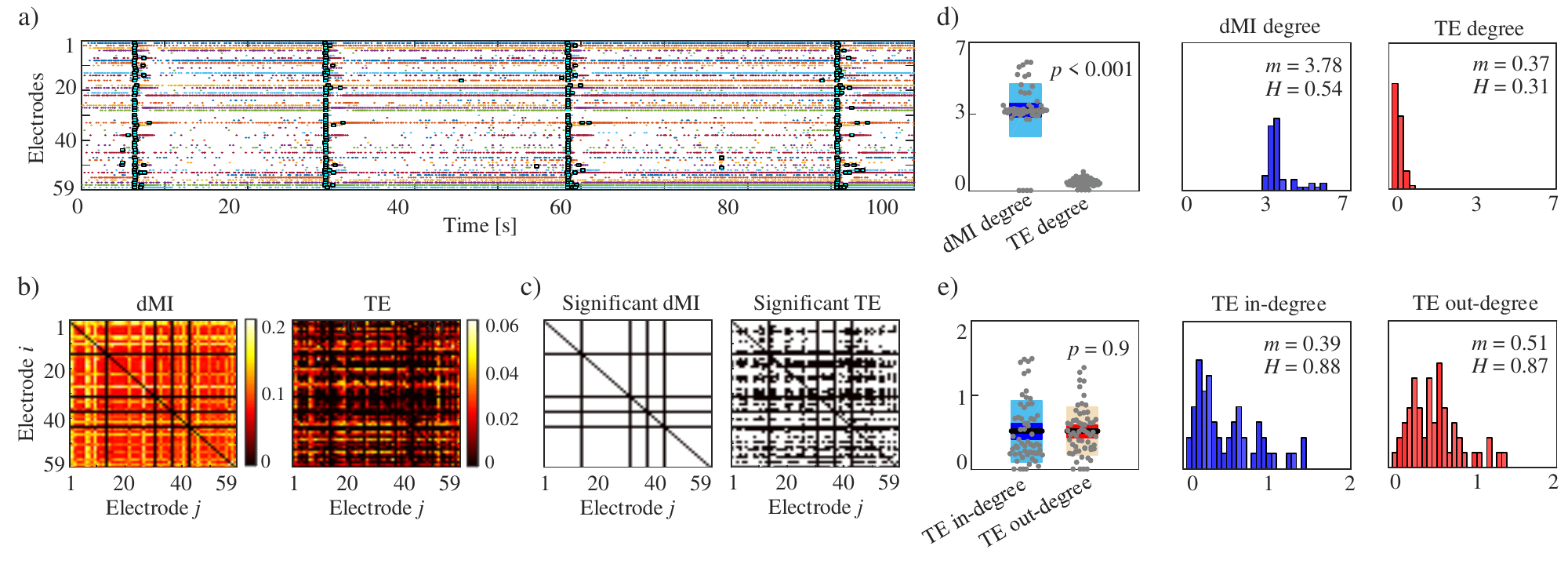}
    \caption{Analysis of dynamic interactions between neuronal spike trains performed for a representative culture in the mature stage (25 days in vitro). (a) Raster plot of the recorded spike trains (dots) for an exemplary portion of the recording, with network bursts detected through temporal clustering denoted by squares. (b) Color-coded matrix representation of the undirected (dMI) and directed (TE) interaction measures computed between each pair of trains for the whole-duration recordings ($\sim 2800$ s). (c) Binary matrix representation of the statistically significant (white) and non-significant (black) dMI and TE values (significance assessed by JODI surrogates). (d,e) Boxplot distribution (with p-value of a paired Wilcoxon test for equal median) and histogram representation (with median ($m$) and entropy ($H$) values) of the weighted dMI degree and TE degree and of the weighted in-degree and out-degree of TE assessed across nodes; note that histograms consider only the active nodes.}
    \label{fig_results1}
\end{figure*}

Since the most prominent feature of the electrical activity of high-density cortical cultures is their propensity to synchronized firing \cite{morin2005investigating, gross1999origins, wong1993transient}, our analysis was focused on the phenomenon of network bursting, i.e. the sporadic synchronous activity across electrodes. In the analyzed preparation, such activity becomes visible after about 10 days in-vitro (DIV) and develops until 30 DIV, while in the beginning cells are electrically quiescent and after 30 DIV they start to degenerate \cite{morin2005investigating}.
Accordingly, we considered 25 cultures analyzed through three stages of maturation, labeled as early ($\sim$ 7 DIV), developing ($\sim$ 15 DIV) and mature ($\sim$ 25 DIV). For each culture and stage, the analysis was performed as depicted in Fig. \ref{fig_results1} and explained in the following.

First, bursts were detected at the level of single electrode through temporal clustering (Fig. \ref{fig_results1}a),
according to an empirical criterion whereby sequences of at least four consecutive spikes were grouped into a burst if all ISIs are lower than a threshold empirically set as the minimum between $1/(4\overline{\lambda})$ and 0.1 s; the minimal ISI was set adaptively, with the value 0.1 s empirically accounting for a worst-case scenario of the timing of axonal propagation among neurons \cite{minati2019connectivity}).
Afterwards, the timing of each burst was identified as the center of mass of the temporal distribution of the spikes within the burst.
We chose to perform this coarse-graining procedure as setting a temporal mesoscale is in this case more representative of a continuous process since the individual spikes are binned by the temporal resolution of the ADC converter. Without this procedure, our analysis performed at the level of individual spikes did not yield significant dMI or TE values; this can be explained considering that spikes outside the bursts are largely stochastic and unrelated to connectivity, while spikes within a burst represent a variety of complex phenomena which cannot be fully disentangled by MEA electrodes each sampling many neurons \cite{minati2019connectivity}, even if complex features are otherwise visible \cite{downes2012emergence,massobrio2015self}.

Consequently, the burst spike trains obtained by temporal clustering (see Fig. \ref{fig_results1}a) were analyzed in pairs estimating the proposed  information-theoretic measures in continuous time to get a symmetric dMI rate matrix and an asymmetric TE rate matrix (Fig. \ref{fig_results1}b).
Estimation was performed setting $N_U=N_X$ and $k_{global}=5$ as in the simulations, and using $l=1$ intervals for history embedding; while the choice of the dimension of embeddings remains crucial in real data analysis, here the investigation of longer uniform embeddings ($l=2, l=3$) did not lead to evident differences (results not shown for brevity).
The statistical significance of each measure (dMI or TE) estimated from a pair of spike trains was assessed generating 100 JODI surrogate trains, using inverse percentiles to derive the probability that the surrogate measure is larger than the original measure, and comparing this probability with a critical significance level set to $\alpha = 0.01/M$ ($M=59$ is the network size); this achieved a Bonferroni correction for the inference of non-isolated network nodes (i.e., for each node,  there is a 0.01 chance that it is connected to at least another node under the null hypothesis of disconnection).
Statistical testing provided a threshold to validate connections in the dMI and TE rate matrices; after binarization of the directed dMI network and of the directed TE network (Fig. \ref{fig_results1}c), we computed the percentages of active nodes and of significant links in each network.
Moreover we studied the distributions of the weighted degree of dMI and TE rates (Fig. \ref{fig_results1}d), as well as the distributions of the weighted in-degree and out-degree of the TE rate (Fig. \ref{fig_results1}e), in terms of statistical comparison (paired Wilcoxon signed-rank test) and of histogram representation (normalized entropy).

\begin{figure*} 
    \centering
    \includegraphics[width=18cm]{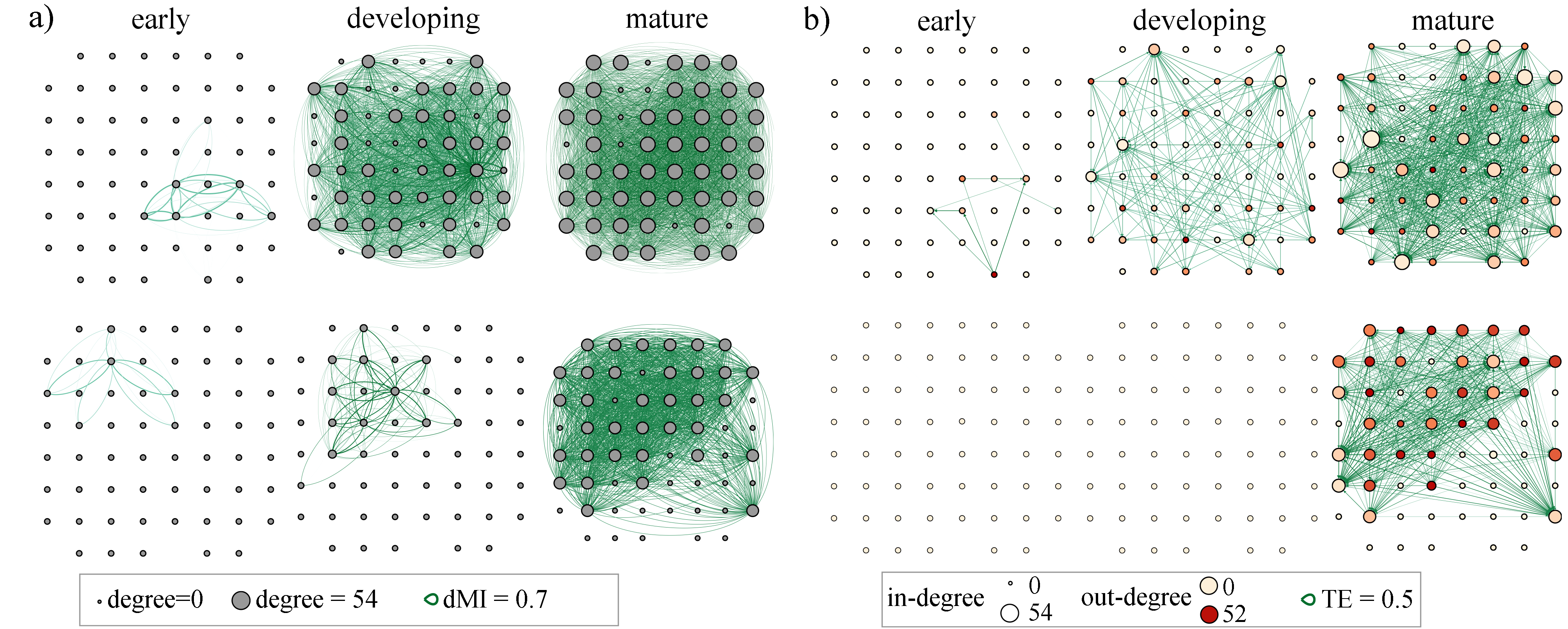}
    \caption{Undirected dMI networks (a) and directed TE networks (b) obtained through information-theoretic analysis of spike trains from two neuronal cultures (rows) considered at different age (early, $\sim$ 7 DIV; developing, $\sim$ 15 DIV; mature, $\sim$ 25 DIV). Green intensity denotes edge strength (dMI or TE magnitude), node size is proportional to dMI degree in (a) and to TE in-degree in (b), node color in maps TE out-degree in (b).}
    \label{fig_results2}
\end{figure*}

The results shown for the culture in Fig. \ref{fig_results1} generalize to the 25 analyzed cultures, as depicted in the network representation of Fig. \ref{fig_results2} for two arbitrarily-chosen but representative cultures studies across stages of maturation, as well as in the complete results reported in the supplementary material.
We found that, in line with previous observations showing increasing network activity and connectivity with the age of the cultures \cite{downes2012emergence,minati2017self,minati2019connectivity}, the number of non-isolated nodes and of significant network links increased moving from the early to the developing and mature stages.
The progression across the three stages was highlighted clearly using dMI (Fig. \ref{fig_results2}a), while TE revealed rich network structures eliciting nodes acting as sources and sinks of information flow in the mature stage (Fig. \ref{fig_results2}b);
the two measures detected a comparable percentage of active nodes, whereas the network connectivity was higher using dMI than TE (Fig. \ref{fig_results2} and Fig. S3a,b).
Moving from the early to the developing and mature stages, the weighted node degree became higher and more dispersed across nodes (increasing median and entropy of the degree distribution, Fig. S3c,d, Fig. S4); upon maturation, the weighted degree derived from dMI was stronger (higher median) and more dispersed across nodes (higher entropy) than the degree derived from TE (Fig. \ref{fig_results1}d and Figs. S3c,e, Fig. S4a,b).
The distributions of the weighted in-degree and out-degree of TE rate became also more skewed and with higher entropy through the maturation process, but did not differ significantly from each other (Fig. S3d,f and Fig. S4c,d).

These results document that both the dMI rate and the TE rate can detect the expected larger involvement of the neuronal units in the establishment of networked functional interactions occurring as the neural cultures spontaneously develop their anatomical connections. Mirroring our simulations, the directed interactions captured by dMI are generally of higher intensity than the undirected interactions captured by TE.
The dMI was better able to describe the gradually raising heterogeneity of the nodal strength also observed in previous studies for these spontaneously growing neuronal networks \cite{downes2012emergence, minati2017self}. In the mature stage however, both dMI and TE highlighted the emergent structural organization of the functional connections between neurons, showing fatter tailed node degree distributions with the appearance of groups of high-strength nodes. The emergence with maturation of a significant proportion of highly connected nodes (hubs) was revealed also in a previous extensive analysis of the firing activity of these cultures, where it was related to the small-world topology of the underlying functional networks \cite{downes2012emergence}. Small-world networks have an architecture which supports efficient information transfer \cite{latora2001efficient}, and the increasing appearance of highly connected nodes in older cultures suggests that such hubs may play a role in organizing the information transfer, possibly acting as sources or sinks for the network activity. This hypothesis is supported in our analysis by the heavy tailed distributions of the in- and out-degree of the TE rate in the mature cultures, where high degrees were observed for a small but non-negligible number of nodes.
Of note, our analysis could not establish the prevalence of sinks or sources of information transfer, as we found a high similarity between the distributions of weighted in- and out-degree of the TE rate.
Such similarity may be explained methodologically: with the rise of synchronized activity, pairwise interactions tend to appear more bidirectional due to the increasing influence of spurious effects (e.g., cascade, common driver) due to processes excluded from the bivariate analysis \cite{smirnov2013spurious,novelli2020inferring}. An extension of the TE rate to multivariate spike trains \cite{shorten2020estimating} is recommended to face this issue and explore more efficiently the topological properties of large-scale and densely connected networks like those emerging from the activity of these neuronal cultures.


\section{Discussion}
Although information-theoretic methods are widely used for the analysis of multivariate time series in computational neuroscience and physiology \cite{lizier2012local,wibral2014directed,porta2015wiener,faes2017information}, their application to spike train data is far less popular. The main reason for this is methodological, being related to the difficulty of a reliable implementation in continuous time of methods intrinsically defined for discrete-time processes like MI, TE, and many other measures of information dynamics. In fact, while a continuous-time formalism can be avoided in the study of processes which are intrinsically occurring in discrete time (e.g., cardiovascular variability series \cite{porta2015wiener,faes2017information}) or can be reasonably represented through sampling techniques (e.g., electromagnetic neural signals \cite{wibral2011transfer,lizier2011multivariate}), it becomes of utmost importance when the information carried by the analyzed process relies on its continuous-time nature, as happens for point processes and for neural spike trains in particular \cite{shorten2020estimating}. In these processes, the analysis based on time discretization - though being common \cite{gourevitch2007evaluating,neymotin2011synaptic,harris2019energy} - is impractical due to issues of estimation bias, data requirement and inability to capture interactions deployed over multiple time scales \cite{shorten2020estimating}. These issues are observed also in our simulations, where we confirm the high bias and strong dependence on data size (number of spikes required) and analysis parameters (temporal bin size and history embedding length) of the discrete-time estimator of the TE rate, and find similar issues for the dMI rate. On the contrary, the adoption of a continuous time formalism implemented with an accurate nearest-neighbor estimator, developed as in \cite{shorten2020estimating} for the TE rate and newly designed for the dMI rate, leads to overcome all shortcomings of the traditional estimates, yielding low-bias, data-efficient and essentially parameter-free measures of directed information flow and neural synchrony. In addition, being based on the non-parametric estimation of information measures, our approach is model-free and thus inherently nonlinear. This stands in contrast to most of the methods explicitly devised to capture directed interactions for point processes, which make use of linear parametric models \cite{kim2011granger, valenza2017instantaneous}.

Our developments build on recent work offering for the first time the possibility to assess in continuous time the directed transfer of information (TE) between event-based data \cite{shorten2020estimating}. This approach, bringing the computational reliability discussed above, opens unique possibilities for the description of brain activities in terms of information flow when such flow is encoded by neural spikes \cite{brown2004multiple}.
Furthermore, following a dominant trend in neuroscience whereby the degree of concurrent firing of neural spike trains is assessed to quantify the general concept of neural synchrony \cite{cutts2014detecting}, we frame into our continuous-time information-theoretic analysis also the evaluation of a symmetric measure of correlation between pairs of trains.
Indeed, besides computing the TE, we define and quantify also the dMI as a dynamic form of mutual information. This measure addresses symmetric interactions that are not captured by the TE, providing the complementary information needed to evaluate thoroughly the whole dynamic interaction between two processes (i.e., the interaction involving different temporal states in the processes, see Fig. 1). Importantly, the undirected dynamic information provided by dMI turns out to be a useful complement to the directional information provided by the TE. In our simulations, we found that dMI can capture better than TE conditions of weakly coupled processes in which the coupling direction does not emerge clearly. These conditions are encountered often in the analysis of neural spike trains, where the synchronous firing of different neural units is not clearly directional and is not consistent in time (e.g., due to the complex properties of neuronal firing and/or to inaccuracy of spike sorting). In the analysis of real-data the dMI rate captured better than the TE rate, in terms of properties of the weighted node degree distribution, the development of the neuronal cultures, showing how network structures containing some densely connected nodes are formed upon maturation. 

In summary, the proposed information-theoretic framework provides principled measures to assess pairwise interactions in point process data in a more robust and flexible way than the discrete-time or parametric approaches previously proposed, and has thus potential to provide new physiological insight into the functional coupling among neural spike trains.
To further improve such potential, future work is envisaged which extends the bivariate approach and allows embracing a fully multivariate perspective on the analysis of spike train data.
The definition of a whole set of measures, hierarchically organized to quantify self-interactions in a spike train (e.g., via information storage \cite{lizier2012local}) as well as pairwise and higher-order interactions among multiple spike trains (e.g., via conditional TE or MI measures \cite{wibral2014directed,faes2015estimating}, or via synergy/redundancy measures \cite{wibral2017quantifying,faes2017information}), will offer the possibility to study generalized network structures where interactions of different orders are represented \cite{battiston2020networks}.
The interest for these enhanced representations of spike train networks extends to other modalities for brain monitoring (e.g. fMRI \cite{tagliazucchi2012criticality,wu2013blind}), and reaches far beyond neuroscience encompassing research fields as diverse as physiology, social systems, seismology and finance \cite{ogata1999seismicity,barbieri2005point,bowsher2007modelling,ver2012information}.

\begin{acknowledgments}
The Authors acknowledge Joseph T. Lizier and David P. Shorten for fruitful discussion.

\end{acknowledgments}



\bibliography{TEMIref}

\end{document}